# Naked Eye Three-dimensional Display System Based on Time-multiplexed Technology

*Ziyang Liu\*, Zekai Chen\*, Changxiong Zheng\*,\*\* Phil Surman\*, Xiao Wei Sun\**

*\*Dept. of Electrical and Electronic Engineering, Southern University of Science and Technology, Shenzhen, Guangdong, China*

*\*\* PengCheng Laboratory, Shenzhen, Guangdong, China.*

**Abstract**

*Our group is developing a multi-user eye-tracked 3D display, an evolution of the single-user eye-tracked 3D display that we have already successfully developed. This display utilizes a slanted lenticular setup, where multiple perspective views are shown across the viewing field. Due to the constraints of the lenticular lens parameters, identical views are repeated across the field, limiting eye tracking to a single user. However, this limitation can be addressed using spatio-temporal multiplexing, where view zone groups are presented sequentially with a high frame rate liquid crystal display (LCD) and driver, in combination with a synchronized directional light emitting diode (LED) array. In this paper, we describe the operation and results of the backlight drive electronics, where a prototype using a white LED illumination matrix, a simplified LCD panel, and a linear Fresnel lens array serves as a test bed.*

## 1. Introduction

As technology advances, the demand for visual experiences among humans continues to increase. Among the most groundbreaking advancements in visual display technology, three-dimensional (3D) displays stand out as a key area of focus for researchers and the display industry. Unlike traditional two-dimensional images, 3D displays cater more effectively to human binocular vision, offering a more accurate and lifelike representation of the physical world. This enhances the overall viewing experience by creating a deeper sense of immersion.

The evolution of technologies such as virtual reality, augmented reality, and mixed reality has played a significant role in driving the development of three-dimensional display technology. However, many mainstream applications are limited by wearable devices, such as AR or VR glasses, resulting in relatively poor user experiences. Consequently, naked eye 3D technology has emerged as the "ultimate way" of information display. [1–4] It requires no additional auxiliary equipment. People can watch three-dimensional images directly with the naked eye, providing an unparalleled visual experience.

The rapid progress in technologies such as the Internet of Things, optical design, computer image processing, and information display has created a conducive environment for the realization of glasses-free three-dimensional display systems. These advancements have paved the way for innovative solutions that enhance user experiences and push the boundaries of visual display technology. [5,6]

Despite the 3D content capabilities of grating system, certain challenges and limitations persist. [7-12] For instance, existing eye-tracking systems are tailored for individual users, presenting a hurdle in ensuring optimal 3D experiences for multiple users simultaneously.

To overcome these challenges, we have introduced time-division multiplexing (TDM) technology alongside the development of new hardware and algorithms. This innovation has enabled the creation of a glasses-free 3D display system specifically designed for multiple users. By effectively managing the viewing experience for different users, this system not only improves user satisfaction but also enhances the overall visual effects.

This study presents an innovative approach to glasses-free 3D display technology, with a focus on delivering a highly immersive and customizable viewing experience. Our experimental methodology employs LCD and LED screens with time-division multiplexing to enable the simultaneous display of personalized 3D content for multiple viewers.

## 2. Configuration

In this part, we will discuss the different components of the three-dimensional imaging system, which comprises five key components: the electronic microprocessor system, the LED (Light-emitting diode) backlight module, the diffuser, the Fresnel Lenses and the LCD (Liquid crystal display) screen.

### 2.1 LED backlight module

In order to realize the naked-eye 3D display system for multiple users, the backlight module needs to be rearranged. To achieve the desired directional lighting properties, we choose individual LED columns rather than a uniform LED backlight. Here, 96 columns of backlight LED light strips are selected and arranged on the back panel.

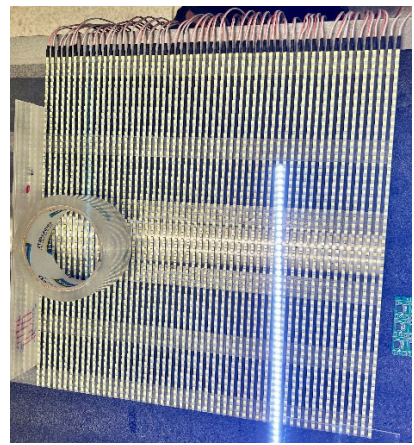

**Figure 1.** Design of LED Backlight Module.

### 2.2 Diffuser

The diffuser is a board attached to the front of the LED plane. Its function is to blur the light emitted by the LED, thereby reducing crosstalk and achieving better display effects.

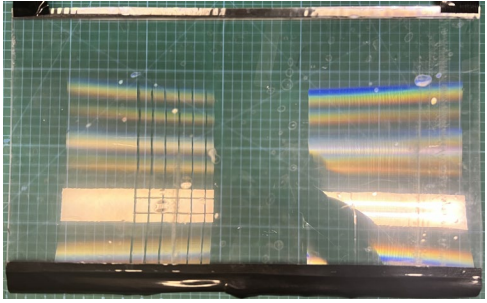

**Figure 2.** Diffuser Layer.

### 2.3 Lenses

The backlight is composed of a series of linear lenses positioned in front of a white LED module. This matrix is arranged into columns, which are selectively activated according to the viewers' eye positions.

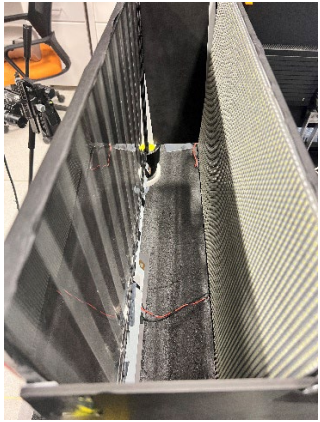

**Figure 3.** A Series of Linear Lenses in front of LED Module.

### 2.4 LCD screen

At the same time, we put a 240Hz LCD display screen in front of it. Through the method of time division multiplexing, we can distribute the 120Hz content to view 1 and the 120Hz content to view 2. Thus, we have completed the construction of a multi-user naked eye 3D display system with TDM.

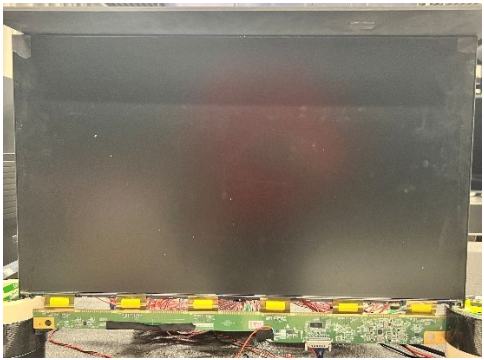

**Figure 4.** LCD Screen.

### 2.5 Electronic microprocessor system

To bring the concept of directional backlighting to realization, we employ the Stm32 microcontroller system—an accessible and efficient platform for innovation. Through programming on the Stm32, we can manipulate each LED column independently, thereby ensuring the time-multiplexed nature of the backlight. This, in turn, enables the time-multiplexed display of distinct viewpoints.

In the following we exhibit the diagram of the signal waveform. The period of the signal is four, known as T1, T2, T3 and T4. The following will present how the microprocessor system works.

This following system is designed to two views, which are the left and right views for the user. At T1, the backlight is turned off, while the LCD refreshes for the left views. At T2, the backlight for the left views is turned on, while LCD does not refresh at all. At T3, the backlight is turned off, while the LCD screen refreshes for the right view. At T4, the backlight for the right view is turned on, while the LCD screen does not refresh at all.

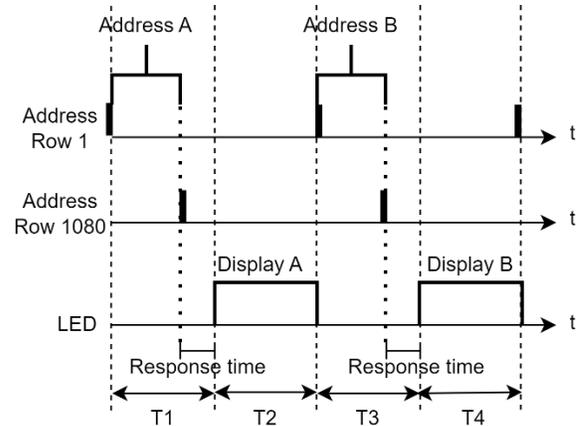

**Figure 5.** Waveform Diagram for signals.

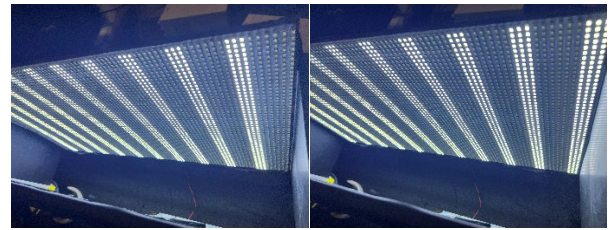

**Figure 6.** LED Time-division Multiplexing Display Effect after Being Controlled by A Microcomputer System

## 3. Principles

In this section, we discuss the fundamental principles of our system.

Exit pupils are formed around the viewers' eyes, where the left and right images are perceived, as illustrated in Fig. 1. In a conventional slanted lenticular display, rays emitted from a point on the display screen can pass through multiple lenses, resulting in conjugate exit pupils, as shown in the simplified diagram of Fig. 1. For clarity, only one pair of conjugate pupils is depicted, though in practice, multiple pairs are likely to form. This configuration works well for single-user displays; however, for multi-viewer scenarios, incorrect images may be seen depending on the relative positions of the viewers.

For each viewer, the left and right exit pupils are simultaneously generated by mapping sub-pixels from the LCD display. These exit pupils are directed only to the intended viewer by controlling the white LEDs in the backlight, based on the output from a multi-user eye position tracker.

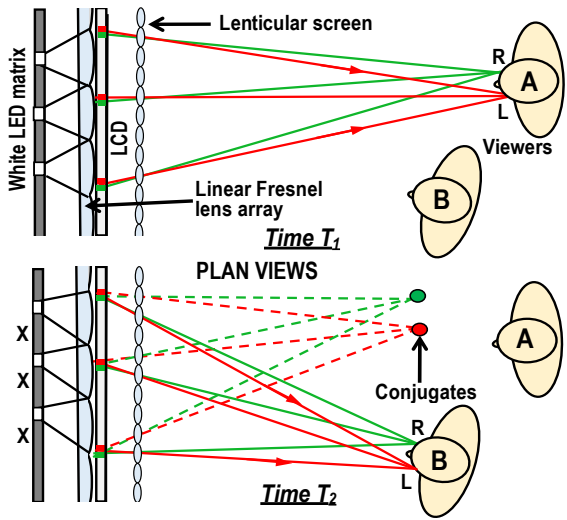

**Figure 7.** Field sequential exit pupils. Viewer A is served at Time T1 and Viewer B at Time T2. Conjugates are not seen by Viewer A as regions X are not illuminated during T2.

The backlight consists of an array of linear lenses placed in front of a white LED matrix. This matrix is organized into columns that are addressed based on the viewers' eye positions. For a system with N viewers, the image on the LCD is refreshed N times per frame, with the backlight switching accordingly for each viewer.

Taking the two-viewer example shown in Fig. 1, the LCD operates at a 120 Hz frame rate, with 2 fields per frame, resulting in an effective LCD refresh rate of 240 Hz. During the first field period, LED columns in the positions indicated in the upper part of Fig. 1 are illuminated. During this period, both eyes of Viewer A see the left and right images simultaneously. On the LCD, images are displayed in adjacent columns of sub-pixels that align with the slant of the lenticular screen. Each lenticular lens corresponds to a column pair on the LCD; note that Fig. 1 shows only a subset of these columns for simplicity.

During the second field, the left and right sub-pixel columns shift, in this example to the right, so that the exit pupils are formed at the eyes of Viewer B. The rays that would have been directed towards Viewer A are inhibited, as no light is emitted from the regions marked 'X' during the second field. In the interval between the fields when the LCD is refreshed, all backlight LEDs are turned off, ensuring that both Viewer A and Viewer B see their respective images simultaneously during the refresh period.

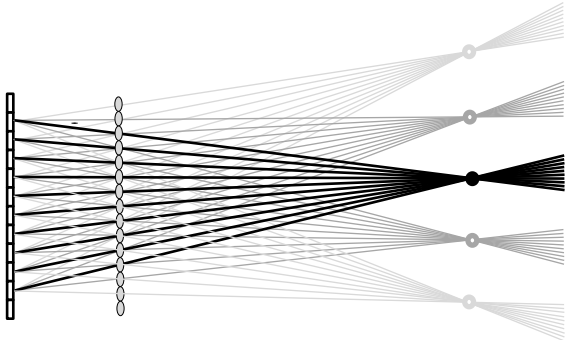

**Figure 8.** Design of LED Backlight module.

This paper will provide a comprehensive overview of the complete display system. However, the primary focus will be on the control and operation of the active backlight LEDs and the LCD.

## 4. Results

The results of the study highlight the exceptional performance of the system in delivering high-quality three-dimensional visual experiences. In this study, we validate our time-division multiplexing (TDM) model using a 27-inch high-definition LED display with a resolution of 2560x1440 (2K). The optimal viewing distance is carefully set at 1 meter to ensure the best possible visual experience for the viewer.

With this configuration, we successfully established a stereoscopic display system. As shown in Fig. 9, three parallelograms of similar width—red, green, and black—are clearly visible. The red parallelogram represents the right-eye view, the green parallelogram corresponds to the left-eye view, and the black parallelogram indicates the gap between the different view zones, which helps minimize crosstalk. The final visual effect aligns with the anticipated outcomes of the time-multiplexed LED system.

The experimental results unequivocally demonstrate that, with the support of TDM technology, our model successfully produces a three-dimensional display effect visible to the naked eye. These findings validate the effectiveness of our model in delivering a high-quality, stereoscopic visual experience.

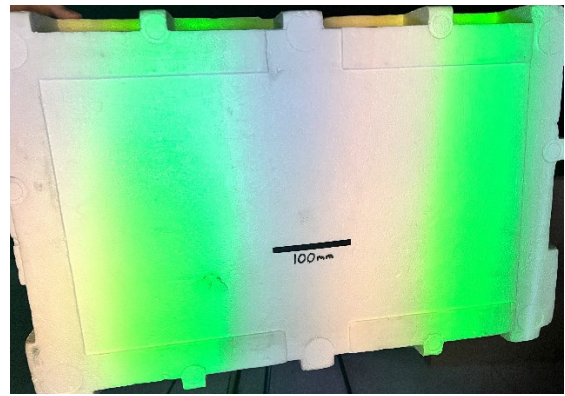

**Figure 9.** The imaging effect at a distance of 1 meter from the display system, where red represents the image formed by the right eye and green represents the image formed by the left eye.

## 5. Future work

In future work, we plan to extend our system from a single-user setup to support multiple users. To achieve this, we will employ an LCD mode with a higher refresh rate to improve display performance. Additionally, we will integrate a depth-sensing camera that will allow us to precisely track the positions of different users within the space. This will enable us to design a new backlight LED mode, tailored to provide optimal views for multiple users simultaneously.

While the current setup works well for a single user, issues may arise when there are multiple viewers. In such cases, incorrect images may be seen depending on the relative positions of the viewers. To address this, each viewer's left and right exit pupils are formed through the mapping of subpixels on the LCD screen. These pupils are directed only to the intended viewer by controlling the backlight, which consists of an array of linear lenses positioned in front of a white LED matrix. For multiple viewers (N viewers), the LCD image will be updated N times per frame, with the backlight adjusted accordingly to ensure correct image presentation for each viewer.

## 6. Conclusion

In this essay, we have proposed an auto-stereoscopic 3D display using time-division multiplexing, which consists of a layer of LED backlight module, a diffuser, a series of lens arrays, a LCD screen and Electronic microprocessor system. The experimental results validate the theoretical principles, showing that our display resolves issues like low resolution, high crosstalk. This system represents an advancement in auto-stereoscopic 3D technology, providing a more immersive 3D viewing experience while maintaining high display quality.

## 7. Acknowledgements

This work was supported by Guangdong University Key Laboratory for Advanced Quantum Dot Displays and Lighting (No. 2017KSYS007), Shenzhen Key Laboratory for Advanced Quantum Dot Displays and Lighting (No. ZDSYS201707281632549), and Shenzhen Development and Reform Commission Project (Grant No. XMHT20220114005):